\documentclass[twocolumn, aps, prl, reprint, superscriptaddress]{revtex4-2}
\usepackage{newtxtext}
\usepackage{amssymb}  % include amsfonts package
\usepackage{amsfonts}  % American Math Society fonts, define \mathbb, \mathfrak 
\usepackage{amsmath}  % multi-lines formulars, \cfrac

\usepackage{amsthm}  % provide a PROOF enviroment
\usepackage{bm}  % bold math and define the vector
\usepackage[version=4]{mhchem} 

\usepackage{bbm}  % hollow math
\usepackage{xcolor}
\usepackage[colorlinks, allcolors=blue]{hyperref}
\usepackage{graphicx}
\usepackage{tabularx}
\usepackage{bbding}
\usepackage{comment}
\usepackage{float}
\usepackage{natbib}
\usepackage{ifthen}
\usepackage{cleveref}
\usepackage{lipsum}
\usepackage{makecell}
\usepackage{braket}
\usepackage[english]{babel}
\usepackage[autostyle, english = american]{csquotes}
\MakeOuterQuote{"}

\theoremstyle{definition}

\theoremstyle{remark}

\usepackage{tikz, datatool, xcolor, ifthen, csvsimple}
\usepackage[framemethod=TikZ]{mdframed}
\usetikzlibrary{arrows, calc, shapes, math, decorations, decorations.markings}
\usetikzlibrary{positioning}
\usetikzlibrary{shapes.geometric}
\usetikzlibrary {arrows.meta}
\usepackage{pgfplots}
\pgfplotsset{compat=newest}
\pgfplotsset{
    colormap={mycm}{rgb255=(225, 225, 225) rgb255=(225, 225, 225)},
    colormap/mycm/.style={
        colormap name=mycm,
    },
}
\usepackage{pgfmath}
\input pgfmath.tex
\tikzset{midarrow/.style={
    decoration={markings, mark=at position 0.55 with {\arrow{>}}},
    postaction={decorate}
}}
\tikzset{midarrowrev/.style={
    decoration={markings, mark=at position 0.45 with {\arrow{<}}},
    postaction={decorate}
}}

\NewDocumentEnvironment{diagram}{O{0.68} O{0.75}}
{
    \begin{tikzpicture}[
        baseline = (X.base),
        every node/.style={scale=#1}, scale=#2
    ]}
    {
    \end{tikzpicture}
}

\definecolor{green}{RGB}{50, 180, 50}
\definecolor{blue}{RGB}{20, 30, 250}
\newcommand{\dobase}[2]{
    \draw (#1,#2) node (X) {$\phantom{X}$};
}

\def\cirrad{0.07}
\def\wtsize{0.07}
\def\shift{0.3}
\NewDocumentCommand{\wt}{m m O{yellow}}{
    \begin{scope}[shift={(#1,#2)}]
        \draw[fill=#3] (0,\wtsize) -- (\wtsize,0) 
        -- (0,-\wtsize) -- (-\wtsize,0) -- cycle;
    \end{scope}
}
\newcommand{\drawTbase}[5]{
    \draw[#5] (#1,#2) to (#1+#4,#2);
    \draw[#5] (#1-#4,#2) to (#1,#2);
    \draw[#5] (#1,#2) to (#1,#2+#4);
    \draw[#5] (#1,#2-#4) to (#1,#2);
    \draw[fill=#3] (#1,#2) circle (\cirrad);
}
\newcommand{\drawTket}[3]{
    \draw[midarrowrev] (#1+\shift,#2+\shift) -- (#1,#2);
    \drawTbase{#1}{#2}{#3}{0.5}{midarrowrev}
}

\begin{document}

\title{Competing pair density wave orders in the square lattice $t$-$J$ model}

\author{Wayne Zheng}
\affiliation{Department of Physics, The Chinese University of Hong Kong, Sha Tin, New Territories, Hong Kong, China}
\author{Zheng-Yuan Yue}
\affiliation{Department of Physics, The Chinese University of Hong Kong, Sha Tin, New Territories, Hong Kong, China}
\author{Jian-Hao Zhang}
\affiliation{Department of Physics and Center for Theory of Quantum Matter, University of Colorado, Boulder, Colorado 80309, USA}
\author{Zheng-Cheng Gu}
\email{zcgu@phy.cuhk.edu.hk}
\affiliation{Department of Physics, The Chinese University of Hong Kong, Sha Tin, New Territories, Hong Kong, China}
\date{\today}

\begin{abstract}
Over the last two decades, the competing orders in high-$T_{c}$ cuprates have been intensely studied, such as pseudogap phase, charge density waves (CDW), and pair density waves (PDW), which are thought to play a crucial role in high-temperature superconductivity. Using the $t$-$J$ model on a square lattice as the simplest model for high-$T_{c}$ cuprates, we employed the fermionic tensor product state (fTPS) method for numerical investigations. Our study revealed new types of PDW states alongside the well-known $d$-wave state and the recently discovered fluctuating PDW state within the low-energy subspace of the $t$-$J$ model. 
%In particular, the so-called plaquette PDW state where the superconducting order parameters change sign on adjoint plaquette has the lowest energy at low doping $\delta <0.05 $. 
We believe that the competition among these states in the underdoped region suggests the potential existence of a fluctuating quantum liquid of PDW states, providing direct evidence for the pseudogap phase's "cheap vortex" scenario. Furthermore, we discuss the potential experimental implication of our discovery.
\end{abstract}

\maketitle

\emph{Introduction ---}
After three decades of intensive investigation since its initial identification in cuprate materials, numerous facets of the high-$T_c$ superconductivity persist as enigmatic puzzles waiting to be solved.
The phenomenon of superconductivity manifests with doping, unveiling a complex phase diagram with various competing orders, including the enigmatic pseudogap phase, the charge density wave (CDW) order, and the recently discovered pair density wave (PDW) order \cite{annurev-conmatphys-031119} characterized by spatially modulating SC order parameter with zero average.
It is generally believed that the effective $t$-$J$ model captures the essence of the high-$T_{c}$ superconductivity in the \ce{CuO_2} plane \cite{BASKARAN1987973, PhysRevB.37.3759, RevModPhys.78.17}.
However, a significant impediment in the study of such strongly correlated quantum many-body systems is the infamous sign problem \cite{PhysRevLett.94.170201, RevModPhys.94.015006}, which thwarts quantum Monte Carlo (QMC) simulations. Fortunately, tensor network algorithms such as the density matrix renormalization group (DMRG), along with its equivalent matrix product states (MPS) for (quasi) 1D systems and 2D tensor product states (TPS) \cite{Orus2014, Orus2019, Banuls2020, RevModPhys.93.045003}, have emerged as robust and promising methods for investigating these systems. Notably, DMRG simulations have validated the emergence of $d$-wave superconductivity (SC) and partially filled stripe/CDW order from the $t$-$J$ model \cite{SCHOLLWOCK201196, science.aal5304, PhysRevLett.127.097003, 2304.03963, 2311.15092, pnas.2109978118, PhysRevLett.127.097002, PhysRevB.108.054505}. 

Recent accumulating experimental evidences indicate the existence of PDW order in cuprate materials, sparking extensive theoretical investigation. 
Scanning tunneling microscopy of both Cooper pairs and single electrons in \ce{Bi_2Sr_2CaCu_2O_{8+x}} has revealed the coexistence of $d$-wave superconductivity and PDW order with $4a_0$ \cite{Hamidian2016, Ruan2018} or $8a_0$ period \cite{Du2020}. 
Additionally, the interaction between PDW and $d$-wave superconductivity leads to lattice rotational symmetry breaking \cite{Chen2022}, which may be related to the observed nematic order in the pseudogap phase of many cuprate materials \cite{Daou2010,Wu2017,Murayama2019}.
In \ce{Bi_2Sr_2CuO_{6+\delta}}, as doping increases, the PDW order emerges from a checkerboard charge pattern at optimal doping \cite{PhysRevX.11.011007}. 
These experiments demonstrate an intimate relationship between PDW, charge density orders, and nematicity. 
Furthermore, PDW order has been discovered in iron-based \cite{Liu2023-kh} and heavy fermion \cite{Gu2023} superconductors, suggesting it is likely a fundamental feature of unconventional superconductivity.
It is proposed that the PDW may hold the key to understanding the mysterious pseudogap phase \cite{PhysRevX.4.031017, Dai2020, Setty2023}, serving as the “parent state” that intertwines the competing orders \cite{Fradkin2015RMP, annurev-conmatphys-031119}.
 
Theoretically, the emergence of pure PDW order has been established in several toy models \cite{Patrick2019,Hong2020, Hong2024}, and previous TPS calculations \cite{Corboz2014} also suggest that this is still possible for the $t-J$ model. In this work, we employ 2D fermionic TPS (fTPS) algorithms to investigate the $t$-$J$ model on the square lattice and explore potential novel PDW orders within the low-energy subspace. By considering unit cell sizes up to $2 \times 2$, we identify two new types of static PDW states characterized by nearly uniform charge density and staggered magnetization, in addition to the fluctuating PDW state recently uncovered in Ref.~\cite{Yue2024}. These states exhibit energy comparable to or lower than the uniform $d$-wave state for $t/J=3.0$ at low dopings.
We believe that these $2a_0$ PDW states may be closely related to the experimentally observed $4a_0$ pattern.

\begin{figure}[tb]
    \centering
    \includegraphics[width=0.43\textwidth]{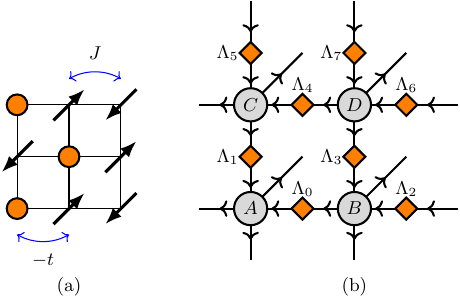}
    \caption{
        (a) Schematic $t$-$J$ model on a square lattice.
        (b) A fTPS comprised of a $2 \times 2$ unit cell on a square lattice, containing four independent fermionic tensors $A, B, C, D$ on lattice sites and eight independent Schmidt fermionic matrices $\Lambda_0, ..., \Lambda_7$ on nearest neighbor bonds.
        Bonds of a tensor corresponding to super vector spaces (or dual spaces) are represented by outgoing (or incoming) arrows. 
    }
    \label{fig:tJ_TN}
\end{figure}

\begin{figure}[tb]
    \centering
    \includegraphics[width=0.42\textwidth]{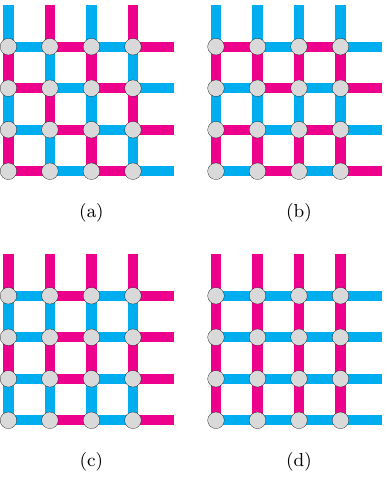}
    \caption{
    Illustration of different uniform-amplitude PDW states obtained from fTPS simulations of the square lattice $t$-$J$ model with $t/J=3.0$.
    (a, b, c) correspond to P0, P1 and P2 in the main text.
    (d) is the conventional already-known $d$-wave state.
    Magenta (or cyan) bonds represent the singlet pairing $\Delta>0$ (or $<0$).
    % Plotted bond width is proportional to the computed $|\Delta_{ij}|$ around the optimal doping $\delta\approx 12\%$.
    }
    \label{fig:tJ_pdw_patterns}
\end{figure}

\emph{Model and methods ---}
The $t$-$J$ model Hamiltonian reads:
%(see Fig. \ref{fig:tJ_TN}(a))
\begin{equation}
    \begin{aligned}
    H &= -t \sum_{\langle{ij}\rangle,\sigma}\left(
        \tilde{c}_{i\sigma}^{\dagger} \tilde{c}_{j\sigma} + h.c.
    \right) + J\sum_{\langle{ij}\rangle}\left(
        \mathbf{S}_{i}\cdot\mathbf{S}_{j}-\frac{n_{i}n_{j}}{4}
    \right),\nonumber
    \end{aligned}
    \label{eq:ham_tJ}
\end{equation}
where 
$
    \tilde{c}_{i\sigma} 
    = c_{i\sigma}(
        1 - c^\dagger_{i\bar{\sigma}}
        c_{i\bar{\sigma}}
    )
$
is the electron operator defined in the no-double occupancy subspace.
$n_{j}\equiv\sum_{\sigma}c_{j\sigma}^{\dagger}c_{j\sigma}$ is the particle number operator on site $j$.
$\langle{ij}\rangle$ denotes a nearest-neighbor (NN) bond.
The spin-1/2 operator is given by 
$
    \mathbf{S}_i = (1/2) \sum_{\alpha,\beta}
    c^\dagger_{i\alpha} \boldsymbol{\sigma}_{\alpha \beta} c_{i\beta}
$, 
where $\boldsymbol{\sigma} = (\sigma^x, \sigma^y, \sigma^z)$ are the Pauli matrices. 

Although a non-uniform strip-like state dominates near a doping of $\delta=1/8$, a previous study indicates that uniform states actually exhibit significantly lower energy at small dopings $\delta<0.05$. In this investigation, our focus lies on the translational-invariant infinite fTPS \cite{PhysRevB.81.165104, PhysRevB.88.115139} within a $2 \times 2$ unit cell, as illustrated in Fig. \ref{fig:tJ_TN}(b):
\begin{equation}
\Ket{\Psi} = \sum_{{s}}\operatorname{fTr} \left(T^s \otimes\Lambda\cdots\right) \Ket{{s}},
\end{equation}
where "$\operatorname{fTr}$" denotes a fermionic contraction adhering to a canonical $\mathbb{Z}_{2}$-graded tensor product isomorphism \cite{Bultinck2018} of all inner virtual bonds in the fTPS. Each bond of a fermionic tensor further divides into two channels characterized by a $\mathbb{Z}_2^{f}$ fermion parity \cite{PhysRevB.88.115139, PhysRevB.95.075108, Bultinck2018, PhysRevB.101.155105, Mortier2024}. The total bond dimension is defined as $D\equiv{D}_{e}+D_{o}$. Each bond possesses a Schmidt weight $\Lambda$.
%which constitutes a rank-$2$ f-tensor (matrix). 
The physical indices are three-dimensional, spanning three orthogonal states: $\ket{0}$, $\ket{\uparrow} = c_{\uparrow}^{\dagger}\ket{0}$, and $\ket{\downarrow} = c_{\downarrow}^{\dagger}\ket{0}$.

Starting from a randomly initialized state $\ket{\Psi}$, we employ imaginary-time evolution to approach the ground state $\ket{\Psi_{0}} \propto\lim_{\tau\rightarrow\infty}e^{-\tau{H}}\ket{\Psi}$. Both the simple and cluster update methods \cite{PhysRevLett.98.070201, PhysRevLett.101.090603, 1110.4362} are utilized; for further details, refer to the Supplementary Material. Doping is regulated by adjusting the chemical potential $\mu$ in the grand canonical ensemble. Our primary focus centers on $t/J=3.0$, as experiments suggest for real materials \cite{RevModPhys.78.17}. In comparison with the $t-J$ model on a honeycomb lattice, we observe strong inhomogeneous tendencies in the square lattice case during practical numerical simulations, aligning with numerous earlier observations of inhomogeneous states in cuprate materials \cite{Tranquada1995, Keimer2015, science.aak9546}. To streamline the investigation and unveil the fundamental physics amidst potential complications, we concentrate primarily on states with uniform charge density and (staggered) magnetization. This is achieved by artificially averaging the bond weights $\Lambda$ during the imaginary time evolution process. To explore all possible local minima under the weighted averaging scheme, we also implement the so-called slave fermion representation, detailed in the Supplementary Material. Notably, inhomogeneous states are found to be nearly degenerate with the uniform states identified in this study, see Supplementary Material for a detailed discussion.

Variational uniform matrix product state (VUMPS) method \cite{PhysRevB.97.045145, PhysRevB.98.235148, PhysRevB.105.195140, PhysRevB.108.035144} is employed to measure these physical observables directly in the thermodynamic limit.
VUMPS is controlled by its own bond dimension $\chi$.
We find that $\chi \gtrsim 4D$ could guarantee a good convergence for local physics observables(refer to the Supplementary Materials for more details; we will fix $\chi=4D$ throughout the whole paper). 
The magnetic order in the obtained state is measured by the staggered magnetization 
$
    M_i = \sqrt{
        \langle S_i^x \rangle^2
        + \langle S_i^y \rangle^2
        + \langle S_i^z \rangle^2
    }
$.
The singlet superconductivity can be directly measured by the real-space singlet pairing on bonds 
$
    \Delta_{ij} = \braket{
        c_{i\uparrow}c_{j\downarrow}
        - c_{i\downarrow}c_{j\uparrow}
    } / \sqrt{2}
$
for the sake of the grand-canonical ensemble.
% $\Delta_{x/y}\equiv\Delta_{\langle{i,i+\hat{x}/\hat{y}}}\rangle$.
Triplet pairing is ignored because its amplitude is essentially zero within our numerical error.

\begin{figure}[tb]
    \centering
    \includegraphics[width=0.45\textwidth]{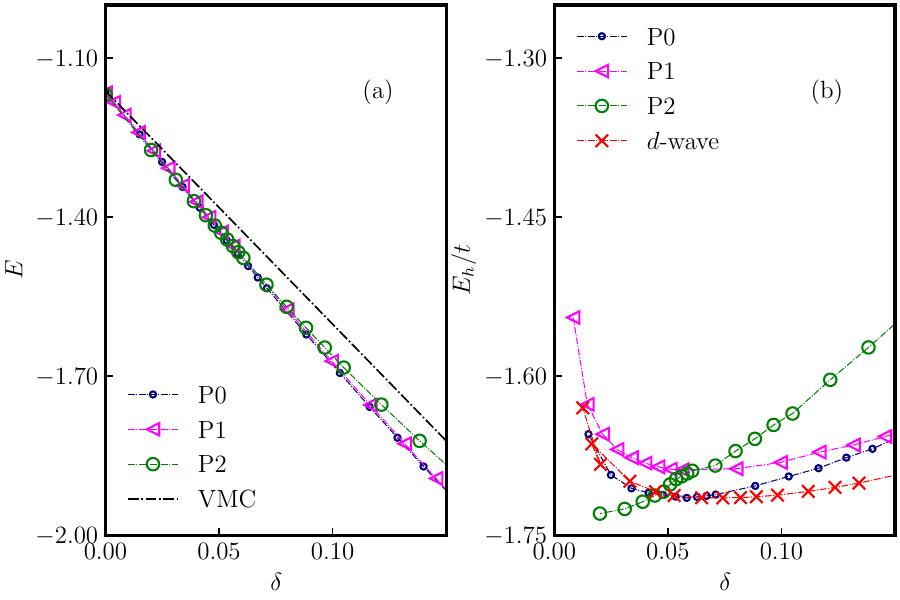}
    \caption{
    (a) Per-site energy $E(\delta)$ of different competing states obtained from $t$-$J$ model with $t/J=3.0$.
    $D=14$.
    The dotted dashed line represents previous VMC results.
    (b) Per-hole energy $E_{h}(\delta)$ of these states. 
    }
    \label{fig:tJ_ene_t30}
\end{figure}

\emph{Competing pair density wave states ---}
The three PDW patterns depicted in Fig. \ref{fig:tJ_pdw_patterns}, along with the $d$-wave state, are ultimately converged from random initial states using our method. Among these patterns, pattern (a) has been previously reported using a smaller fTPS unit cell \cite{Yue2024}. Through weight averaging, the produced states exhibit uniformity in charge density, magnetization $M_i$, and the magnitude of NN singlet pairing $|\Delta_{ij}|$. As seen in Fig. \ref{fig:tJ_ene_t30}, it is evident that the per-site energies of these competitive states are nearly degenerate and notably lower than the previous variational Monte Carlo outcomes at finite dopings \cite{PhysRevB.70.104503, PhysRevB.81.165104}. For $D=14$, up to our computational resource limit, the extrapolated energy at half-filling amounts to $\tilde{E}_0 \approx -1.166(1)$, slightly surpassing by about $0.3 \% $ the best QMC result of $E_0=-1.1694$ for the Heisenberg model \cite{PhysRevB.56.11678}. Pattern P0 can be identified as a pure PDW state, where $\Delta_{ij}$ on NN bonds carries a crystal momentum of $\left(\pi, \pi\right)$,  %Macroscopically, there is an absence of superconductivity in the P0 state since the average of singlet pairs vanishes. 
while P1 and P2 states can be regarded as a mixing of PDW states with different crystal momenta.

\begin{comment}
\begin{Fig.}[tb]
    \centering
    \includegraphics[width=0.48\textwidth]{tJ_vumps_PDWs_D14_chi56.pdf}
    \caption{
    Different PDW patterns with $D=14$ measured with boundary MPS bond dimension $\chi = 4D$.
    (a) Doping $\delta$ as a function of chemical potential $\mu$.
    (b) Ground state energies $E(\delta)$ as a function of doping $\delta$.
    Dashed dotted line denotes the energies from previous VMC.
    (c)  Magnetization magnitude $M(\delta)$.
    (d) Singlet pairings' magnitude $\vert\Delta\vert(\delta)$.
    }
    \label{fig:tJ_D14}
\end{Fig.}

\begin{Fig.}[tb]
    \centering
    \includegraphics[width=0.48\textwidth]{tJ_vumps_PDWs_Dscaling_ene.pdf}
    \caption{
    Estimation of ground state energies as $D \to \infty$ of various PDW states from a $1/D$ extrapolation.
    (a, b, c, d, e) correspond to P0, P1, P2, P3, P4.
    Each $E(\delta)$ is fitted with a second-order polynomial and plotted as a dashed line with the same color.
    Insets are extrapolation examples with $\delta=0.125$.
    }
    \label{fig:tJ_Dscaling_ene}
\end{Fig.}
\end{comment}

\begin{figure}[H]
    \centering
    \includegraphics[width=0.52\textwidth]{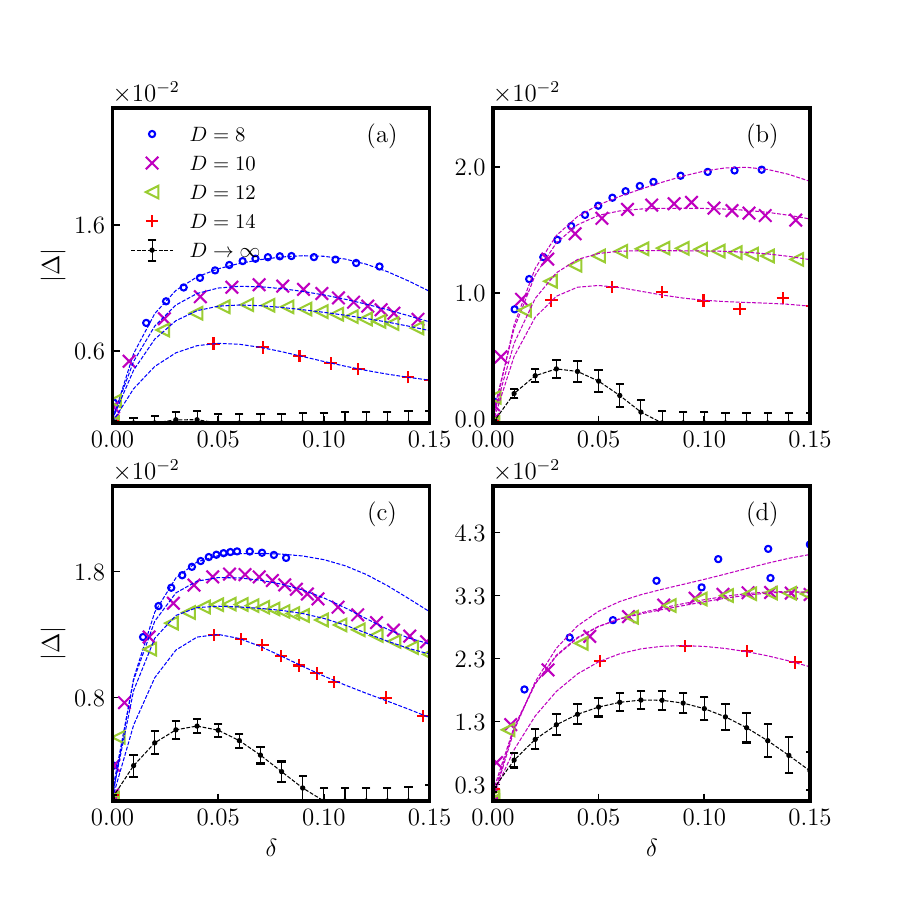}
    \caption{
    Estimation of singlet pairing magnitudes $\vert\Delta\vert(\delta)$ as $D\to\infty$ for competing states in $t$-$J$ model from a $1/D$ extrapolation.
    $t/J=3.0$.
    (a, b, c, d) correspond to P0, P1, P2 and $d$-wave, respectively.
    % Lower Fig. in the second row illustrate $t=2.5$. 
    Each $\vert\Delta\vert(\delta)$ is fitted by a sixth-order polynomial and plotted as a dashed line with the same color.
    %Insets are extrapolation examples with $\delta=0.125$.
    }
    \label{fig:tJ_Dscaling_sp}
\end{figure}

\begin{figure}[tb]
    \centering
    \includegraphics[width=0.52\textwidth]{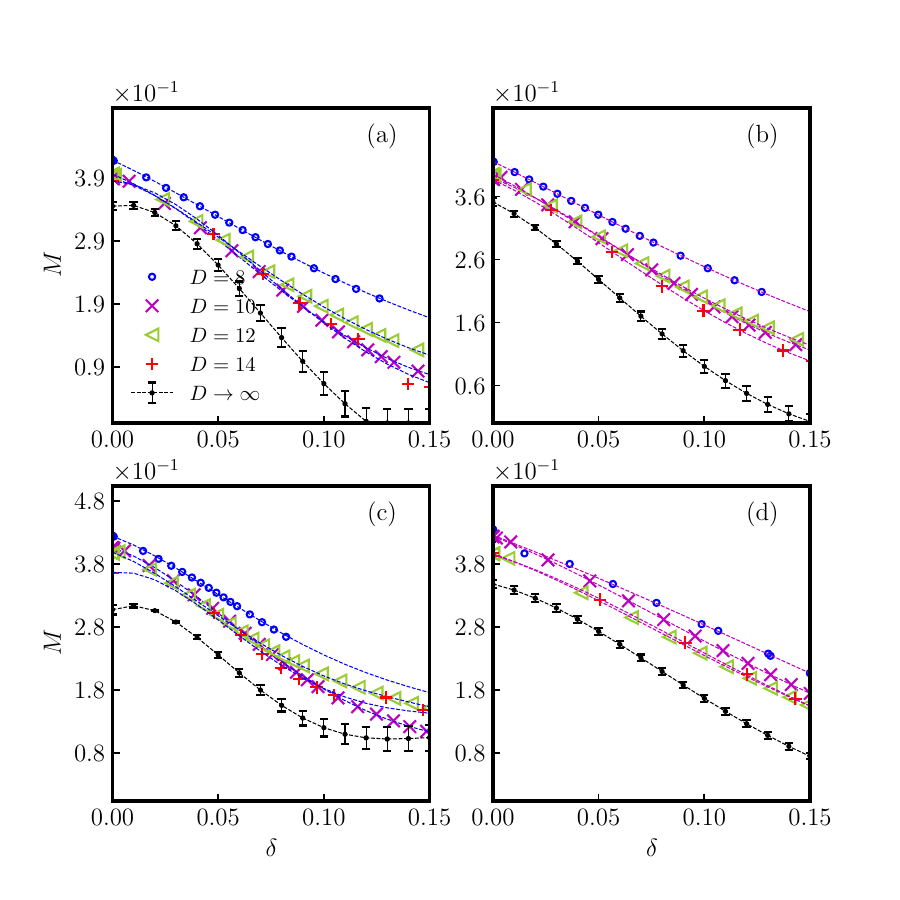}
    \caption{
    Estimation of magnetization $M(\delta)$ as $D \to \infty$ of competing states in $t$-$J$ model from a $1/D$ extrapolation.
    $t/J=3.0$.
    (a, b, c, d) correspond to P0, P1, P2 and $d$-wave state, respectively.
    Each $M(\delta)$ is fitted by a fourth-degree polynomial and plotted as a dashed line with the same color.
    %Insets are extrapolation examples with $\delta=0.125$.
    }
    \label{fig:tJ_Dscaling_mag}
\end{figure}

To further distinguish the nearly degenerate energies of these states, we plot the energy per-hole \cite{PhysRevB.84.041108} defined by 
$E_h(\delta) \equiv [E(\delta)-E_0] / \delta$, which is a re-scaled energy of per-site, 
in comparison with the uniform $d$-wave state reported before \cite{PhysRevLett.113.046402}.
As shown in Fig. \ref{fig:tJ_ene_t30} for $D=14$, we can see that their energies are still very close especially at low dopings.
In this sense, for a doped Mott insulator, it is revealed that the low energy ground states manifold could be highly degenerate and competing in terms of many kinds of real-space paring states including PDWs and $d$-wave.
Particularly, P2 state seems to have the lowest energy at low doping $\delta<0.05$.
We realize that the fruitfulness of the high-$T_{c}$ superconductor phase diagram could be directly related to this microscopic complexity.
% P2 pattern has a higher energy in the overdoped regime.
Next, we make extrapolated estimations of singlet pairing's magnitude and magnetization for these PDW states as $D \to \infty$, which are shown in Fig. \ref{fig:tJ_Dscaling_sp} and Fig. \ref{fig:tJ_Dscaling_mag}.
In Fig. \ref{fig:tJ_Dscaling_sp} we can see that, the plaquette P2 state, possesses the strongest pairing amplitude among PDW states in the underdoped regime.
We note that the paring amplitude in P1, P2 and $d$-wave states can survive in the limit $D\to\infty$.
% However, it seems that our methods seem having a limitation for the estimation of magnetization, especially for P2 pattern as shown in Fig. \ref{fig:tJ_Dscaling_mag}.
As shown in Fig. \ref{fig:tJ_Dscaling_mag}, the magnetization decreases as dopings increase, 
which, however, only coincides with experiments qualitatively but not quantitatively.
In the real case, antiferromagnetism is suppressed much quicker in the phase diagram of copper oxide \cite{Keimer2015}.
We ascribe this to other secondary effects and inaccuracy of simple/cluster update method. In our future study, we will also consider more realistic models and implement the full update method to improve this results.
%Real materials certainly cannot be thoroughly described by a pure $t$-$J$ model.

\emph{Effect of next nearest neighbor hopping and t/J ratio ---}
To better describe realistic cuprates, next-nearest neighbor (NNN) hopping terms $-t' \sum_{\braket{\braket{ij}}} (c^\dagger_{i\sigma} c_{i\sigma} + h.c.)$ should be added to the Hamiltonian, where $t' < 0$ in the hole-doped case, and $t' > 0$ in the electron-doped case \cite{Eskes1989,Tohyama1994}. 
To study how it affects the PDW states, we measure the hopping $T_{ij} = \sum_\sigma \braket{c^\dagger_{i\sigma} c_{i\sigma}}$ on NNN bonds. 
The results are shown in Fig. \ref{fig:hopping-nnn}. 
In the $d$-wave state, $T_{ij} > 0$ on NNN bonds, which means that a positive $t'$ can further lower its energy. This is consistent with DMRG calculations, which found that $t'>0$ stabilizes the $d$-wave superconductivity \cite{Jiang2021, Jiang2022, Lu2024}. 
In contrast, for the P1 state, $T_{ij} < 0$ on the $d^1_{ab}$ bonds. Although $T_{ij} > 0$ on the $d^2_{ab}$ bonds, it is smaller than $|T_{ij}|$ on $d^1_{ab}$. With the same $t'$ on the NNN bonds, the energy can be lowered with a negative $t'$ instead. 
Similarly, the energy of the other two PDW states P1 and P2 can also be lowered with a negative $t'$. 
Therefore, in actual hole-doped cuprates with $t' < 0$, the PDW states should become more stable than the $d$-wave state at low doping. Finally, it is worth noting that while the pairing amplitude in P2 displays a $C_4$ symmetry, the NNN hopping term (very similar to the fluctuating PDW state P1) also explicitly breaks this symmetry along the diagonal direction.

Different coupling strengths, with $t/J=2.5$ and $t/J=5.0$, exhibiting similar PDW patterns, are detailed in the Supplementary Materials. Similar to the previously discovered fluctuating PDW state, a bigger $t/J$ ratio tends to stabilize PDW states, while a smaller ratio favors the uniform d-wave state at low doping levels. While cluster update may not be the most accurate method for tensor network simulation, our unpublished work utilizing the full update algorithm demonstrates no qualitative changes in the results. Specifically, a bigger $t/J$ ratio is observed to amplify quantum fluctuation and stabilize PDW states. 
%In particular, in the strong coupling limit, P1 exhibits the lowest energy.

\begin{figure}[tb]
    \centering
    \includegraphics[width=0.25\textwidth]{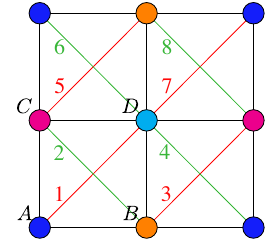}
    \caption{Non-equivalent next nearest neighbor (NNN) bonds in the TPS with a $2 \times 2$ unit cell, labeled as 1 to 8.}
    \label{fig:nnn-bonds}
\end{figure}

\begin{figure}[tb]
    \centering
    \includegraphics[width=0.45\textwidth]{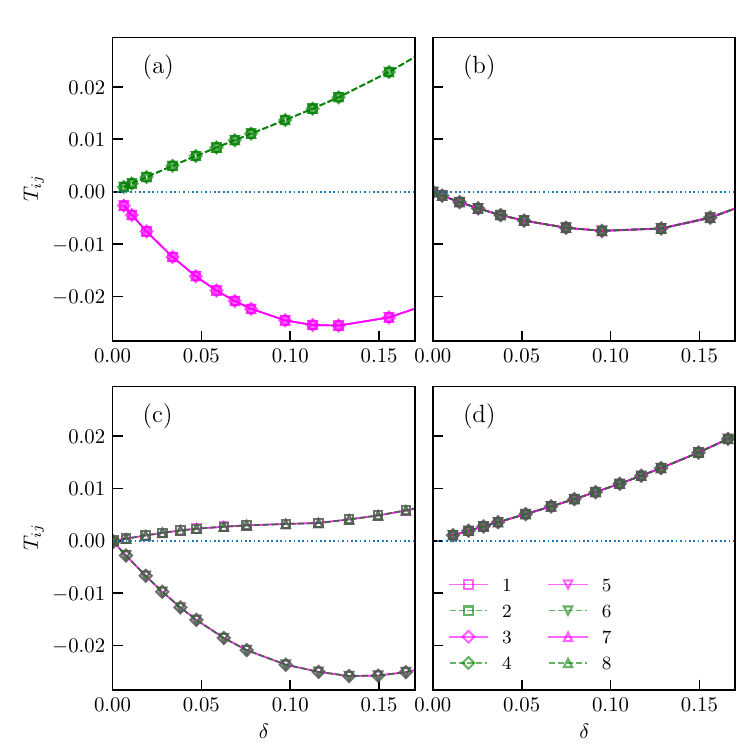}
    \caption{
        Hopping $T_{ij} = \sum_\sigma \braket{c^\dagger_{i\sigma} c_{i\sigma}}$ on NNN bonds (Fig. \ref{fig:nnn-bonds}) in the three PDW states and the $d$-wave state at $t/J = 3.0$.
        $D=10$ and $\chi = 32$. 
        (a, b, c, d) correspond to P0, P1, P2 and $d$-wave. 
        For P1, P2 and $d$-wave, the curves for bonds $1,3,5,7$ overlap with the curves for bonds $2,4,6,8$, respectively (the gray color results from the overlap of green and magenta). 
    }
    \label{fig:hopping-nnn}
\end{figure}

\emph{Conclusion and discussion ---}
In conclusion, by employing a 2D fTPS method, we numerically identify various kinds of nearly degenerate PDW states in the ground states subspace of $t$-$J$ model in addition to the well-known $d$-wave state.
Our findings strongly suggest that the ground state manifold of the $t$-$J$ model exhibits significant degeneracy and complexity. Although globally distinct, P0, P1 and P2 share a common local trait: the superconducting order parameter changes sign upon lattice translation. Thus, we can propose a novel concept—the \emph{PDW liquid (PDWL) } state $\ket{PDWL}=\sum_{s}\alpha_{s} \ket{P_s}$—which comprises a superposition of nearly degenerate PDW patterns $\ket{P_s}$ characterized by maximal sign modulations on neighboring bonds, akin to the three PDW patterns we have identified within a $2\times2$ unit cell.
The PDWL state, featuring a finite paring amplitude but lacking macroscopic superconductivity, emerges as a potential candidate for the pseudogap phase prevalent across the phase diagram of various cuprate materials. This innovative state offers insights into the intricate behavior of doped Mott insulators, shedding light on the diverse phenomena observed in high-temperature superconductors \cite{Corson1999, science.1176369, Ye2023}. 
Certainly, realistic materials introduce a myriad of complexities, including NNN hopping and disorders, which can play a role in stabilizing specific PDW states observed in experimental settings. Recent experimental advancements, particularly in resonant inelastic X-ray scattering (RIXS), have unveiled intriguing insights. These studies reveal the remarkable robustness of local charge modulations with an approximate wave vector of $(\pi/2,\pi/2)$ across a broad range of doping concentrations, persisting up to temperatures around $T\sim 300K$. In contrast, long-range charge order is distinctly evident only up to temperatures of approximately $T\sim 60K$ \cite{Chang_2012, Peng_2018, Arpaia_2019}, underscoring the significance of PDW phenomena in cuprates.
The emergence of PDW states is closely linked to the pseudogap phenomena, with experimental evidence indicating the potential existence of localized pre-pairing singlets in lightly doped Mott insulator phases \cite{Kohsaka2008, 2309.09260}. These findings suggest the emergence of PDW states from the antiferromagnetic phase, independent of Fermi-surface nesting considerations. Furthermore, experimental data strongly suggest that $(\pi/2, \pi/2)$ modulated $4a_{0}$ ($a_{0}$ denotes the lattice constant) checkerboard patterns play a fundamental role in cuprates \cite{Hamidian2016, PhysRevX.11.011007, Ye2023}. Interestingly, the P2 pattern we have observed exhibits a striking resemblance to the $C_4$ symmetric checkerboard pattern in terms of pairing amplitudes. This similarity hints at a hidden $C_4$ symmetry breaking along diagonal directions, presenting a compelling opportunity for future experimental investigations.
%to probe and quantify this symmetry breaking phenomenon.
Of course, due to the restrictions of our simulations to a $2\times{2}$ unit cell, only $(\pi, \pi)$ modulations are currently feasible. Future endeavors aim to address this limitation by expanding unit cells. 
%to encompass these crucial $(\pi/2, \pi/2)$ modulations in our investigations.

\emph{Acknowledgments ---}
WZ would like to thank valuable discussions with Wei Zhang, Shusen Ye, Wen-Yuan Liu, and M. P. A. Fisher.
JHZ thanks Yayu Wang and Zheng-Yu Weng for their insightful discussions.
The authors all thank Zheng-Tao Xu for help on the fermionic VUMPS algorithm. This work is supported by funding from Hong Kong's Research Grants Council (CRF C7012-21GF and GRF No. 14302021).

\appendix

\section{Cluster update method for fermionic TPS}

\begin{figure}[tb]
    \centering
    \includegraphics[width=0.42\textwidth]{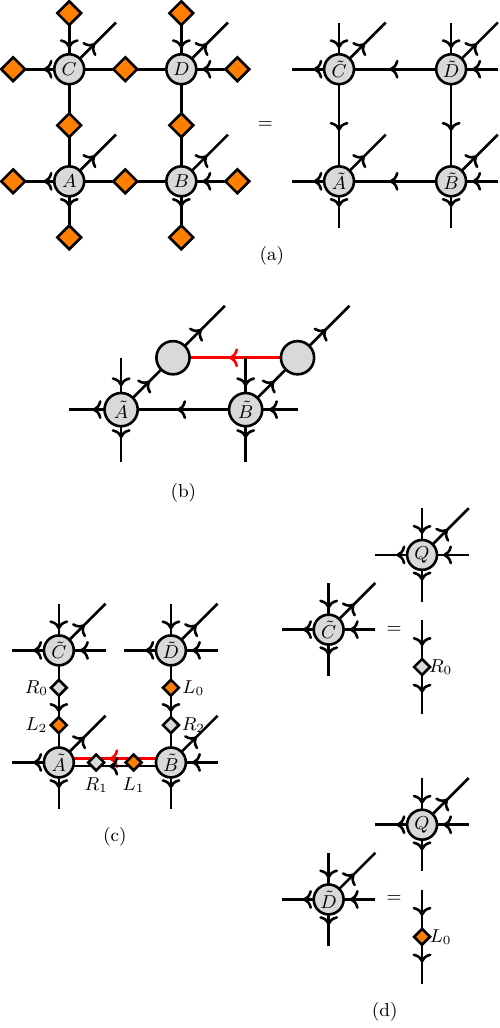}
    \caption{
    \label{fig:cluster_update_method}
    Illustration of the cluster update method for fTPS.
    (a) The whole unit cell and related weight matrices (denoted by large squares) are considered as a cluster.
    They can be contracted into merged tensors.
    (b) A time evolution fMPO operates on $AB$ bond.
    (c) The new cluster is re-formulated as an open fMPS.
    (d) The first $R_{0}$ and $L_{0}$ residual matrices (denoted by small squares) are obtained by QR factorizations from the open ends.
    }
\end{figure}

In this section, we explain some details of the cluster update method~\cite{1110.4362, PhysRevB.94.075143} used in the main text.
We take the $AB$ bond of a $2\times{2}$ unit cell as a representative.
Firstly, as shown in Fig.~\ref{fig:cluster_update_method}(a), to include more environment effects and allow more quantum fluctuations, the cluster consists of all the fermionic tensors in the unit cell and related bond weight matrices (denoted by large squares).
We absorb all bond weights and obtain the so-called \emph{merged tensors} $\tilde{A}, \tilde{B}, \tilde{C}, \tilde{D}$.
Secondly, as shown in Fig.~\ref{fig:cluster_update_method}(b), we apply an imaginary time evolution operator $e^{-\epsilon{H}_{b}}$ onto $\tilde{A}\tilde{B}$.
$e^{-\epsilon{H}_{b}}$ is also written as a 2-site fermionic matrix product operator (fMPO).
Its internal bond dimension is denoted by $\rho$.
After this step, the dimension of the $\tilde{A}\tilde{B}$ bond is increased to $D^{2}\rho$ as illustrated in Fig.~\ref{fig:cluster_update_method}(c).
Our strategy is to build some projectors to compress this increased bond dimension back to $D^{2}$ as good as possible at each imaginary time evolution.
Otherwise, it will increase exponentially.
Thirdly, as illustrated in Fig.~\ref{fig:cluster_update_method}(c), the cluster is regarded as a fermionic MPS with open boundary condition, since only open MPS possesses a canonical form.
Closed loops need further considerations~\cite{PhysRevB.98.085155}.
Next, we perform successive QR and LQ factorizations to obtain the \emph{residual matrices} $R_i$ and $L_i$ (represented by small squares; $i = 0,1,2$) on each bond of the cluster (see Fig.~\ref{fig:cluster_update_method}(d)):
\begin{equation}
\begin{gathered}
    \tilde{C} = Q_0 R_0, \quad
    R_0 \tilde{A} = Q_1 R_1, \quad
    R_1 \tilde{B} = Q_2 R_2;
    \\
    \tilde{D} = L_0 Q'_0, \quad
    \tilde{B} L_0 = L_1 Q'_1, \quad
    \tilde{A} L_1 = L_2 Q'_2.
\end{gathered}
\end{equation}
Next, we place an identity operator on each bond written with these residual matrices:
\begin{equation}
    1 = R^{-1} R L L^{-1}
\end{equation}
and we can perform a SVD in terms of
\begin{equation}
    R L
    = U \Lambda {V} 
    \simeq \bar{U} \bar{\Lambda} \bar{V},
\end{equation}
where $\simeq$ means the singular value spectrum has been truncated to the original dimension of the bond. 
Then the two projectors used to truncate each bond in the cluster are constructed as
\begin{align}
    P^{R}
    &= R^{-1} U
    = L {V}^{\dagger} {\Lambda}^{-1}
    \simeq L \bar{V}^{\dagger} \bar{\Lambda}^{-1}, 
    \\
    P^{L}
    &= V {L}^{-1}
    = {\Lambda}^{-1} {U}^{\dagger} R
    \simeq \bar{\Lambda}^{-1} 
    \bar{U}^{\dagger} R.
\end{align}
Other bonds in the unit cell are updated similarly.

In practice, the time evolution step $\epsilon$ is gradually decreased from $10^{-2}$ to $10^{-6}$.
The algorithm is considered as converged when the change in the Schmidt weights between two update steps $||\Lambda_{\text{new}}-\Lambda_{\text{old}}||<10^{-10}$.
Our code is available at \cite{wei2024}.

\section{$t$-$J$ model in the slave-fermion representation}

In the main text, the fTPS physical index is $(1+2)$-dimensional, where the even subspace is spanned by the vacuum state $\ket{0}$, and the odd subspace is spanned by the spin-up $\ket{\uparrow} = c^\dagger_\uparrow \ket{0}$ and spin-down $\ket{\downarrow} = c^\dagger_\downarrow \ket{0}$ states. 
However, we can equally use a $(2+1)$-dimensional physical index from the slave-fermion point of view \cite{Yoshioka1989}, which decomposes the electron operator as
\begin{equation}
    \tilde{c}_{j\sigma}=h_{j}^{\dagger} b_{j\sigma}, 
    \ \ \text{subject to} \ \ 
    \sum_{\sigma}b_{j\sigma}^{\dagger}b_{j\sigma}
    +h_{j}^{\dagger}h_{j}=1,
\end{equation}
where $h_{j}$ is a spinless fermion (holon) carrying the charge, and $b_{j\sigma}$ is a boson (spinon) carrying the spin. 
Then the even subspace is spanned by the spinon states
$\ket{\uparrow} = b^\dagger_\uparrow \ket{0}$, 
$\ket{\downarrow} = b^\dagger_\downarrow \ket{0}$, 
while the odd subspace is spanned by the holon state
$\ket{o} = h^\dagger \ket{0}$. 
For an fTPS with $2 \times 2$ unit cell, these two choices can be related by a gauge transformation that preserves the parity of each local tensor, and does not affect the expectation values of physical observables. We construct this transformation below. 

First, any single-site operator $\hat{O}$ acting on the physical index in the two representations (denoted as $\hat{O}_f$ in the original electron representation, and $\hat{O}_b$ in the slave-fermion representation) are related by $\hat{O}_b = U_p \hat{O}_f U^\dagger_p$, or graphically
\begin{equation}
    \begin{diagram}[0.8][1.4]
        \dobase{0}{0} 
        \foreach \y in {-0.5,0}
        {\draw[midarrow] (0,\y) -- ++(0,0.5);}
        \draw[fill=white] (0,0) circle (\cirrad);
        \node[anchor=east] at (-0.1,0) {$\hat{O}_b$};
    \end{diagram}
    = \begin{diagram}[0.8][1.4]
        \dobase{0}{0} 
        \foreach \y in {-1,-0.5,0,0.5}
        {\draw[midarrow] (0,\y) -- ++(0,0.5);}
        \draw[fill=white] (0,0) circle (\cirrad);
        \foreach \y in {-0.5,0.5} {\wt{0}{\y}}
        \node[anchor=east] at (-0.1,0.5) {$U_p$};
        \node[anchor=east] at (-0.1,0) {$\hat{O}_f$};
        \node[anchor=east] at (-0.1,-0.5) {$U^\dagger_p$};
    \end{diagram},
\end{equation}
where $U_p$ is an odd-parity tensor
\begin{equation}
    \begin{diagram}[0.8][1.4]
        \dobase{0}{0} 
        \foreach \y in {-0.5,0}
        {\draw[midarrow] (0,\y) -- ++(0,0.5);}
        \wt{0}{0}
        \node[anchor=east] at (-0.1,0) {$U_p$};
        \node[anchor=south] at (0,0.5) {$\mu$};
        \node[anchor=north] at (0,-0.5) {$\nu$};
    \end{diagram} = \left(\begin{array}{c|cc}
        0 & 1 & 0 \\
        0 & 0 & 1 \\ \hline
        1 & 0 & 0
    \end{array}\right)_{\mu \nu} \ket{\mu} \bra{\nu},
\end{equation}
(lines in the matrix separate the even and the odd subspaces) which is unitary ($U^\dagger_p U_p = U_p U^\dagger_p = 1$), or graphically
\begin{equation}
    \begin{diagram}[0.8][1.4]
        \dobase{0}{0.25} 
        \foreach \y in {-0.5,0,0.5}
        {\draw[midarrow] (0,\y) -- ++(0,0.5);}
        \wt{0}{0} \wt{0}{0.5}
        \node[anchor=east] at (-0.1,0.5) {$U^\dagger_p$};
        \node[anchor=east] at (-0.1,0) {$U_p$};
        \node[anchor=south] at (0,1) {$\mu$};
        \node[anchor=north] at (0,-0.5) {$\nu$};
    \end{diagram} = \begin{diagram}[0.8][1.4]
        \dobase{0}{0.25} 
        \foreach \y in {-0.5,0,0.5}
        {\draw[midarrow] (0,\y) -- ++(0,0.5);}
        \wt{0}{0} \wt{0}{0.5}
        \node[anchor=east] at (-0.1,0.5) {$U_p$};
        \node[anchor=east] at (-0.1,0) {$U^\dagger_p$};
        \node[anchor=south] at (0,1) {$\mu$};
        \node[anchor=north] at (0,-0.5) {$\nu$};
    \end{diagram}
    = \delta_{\mu \nu} \ket{\mu} \bra{\nu}. 
\end{equation}
Suppose that we have absorbed the bond weights $\Lambda_a$ ($a = 0,...,7$) into the site tensors $A, ..., D$, so that the fTPS in the original electron representation takes a simpler form, 
\begingroup
\newcommand{\drawlatket}{
    \foreach \x in {0,1}
    \foreach \y in {0,1} {
        \drawTket{\x}{\y}{blue}
    }
    \foreach \y in {0,1} {
        \drawTket{0.5}{\y}{orange}
    }
    \foreach \x in {0,1} {
        \drawTket{\x}{0.5}{magenta}
    }
    \drawTket{0.5}{0.5}{cyan}
}
\begin{equation}
    \ket{\psi_f} = \begin{diagram}[0.8][1.8]
        \dobase{0}{0.5} 
        \drawlatket
        \node[anchor=north west] at (0,0) {$T^f_A$};
        \node[anchor=north west] at (0.5,0) {$T^f_B$};
        \node[anchor=north west] at (0,0.5) {$T^f_C$};
        \node[anchor=north west] at (0.5,0.5) {$T^f_D$};
    \end{diagram}.
\end{equation}
\endgroup
Then we transform $T^f_A, ..., T^f_D$ to
\begin{equation}
\begin{aligned}
    \begin{diagram}[0.8][1.2]
        \dobase{0}{0} \drawTket{0}{0}{magenta}
        \node[anchor=south east] at (0,0) {$T^b_C$};
    \end{diagram} &= \begin{diagram}[0.8][1.2]
        \dobase{0}{0} \drawTket{0}{0}{magenta}
        \node[anchor=south east] at (0,0) {$T^f_C$};
        \draw[midarrow] (\shift,\shift) -- ++(\shift,\shift);
        \wt{\shift}{\shift}
        \node[anchor=south] at (\shift,\shift) {$U_p$};
        \draw[midarrow] (-0.5,0) to (-1,0);
        \wt{-0.5}{0}[lime]
        \node[anchor=north] at (-0.5,-0.05) {$U^{DC\dagger}_v$};
    \end{diagram},
    &
    \begin{diagram}[0.8][1.2]
        \dobase{0}{0} \drawTket{0}{0}{cyan}
        \node[anchor=south east] at (0,0) {$T^b_D$};
    \end{diagram} &= \begin{diagram}[0.8][1.2]
        \dobase{0}{0} \drawTket{0}{0}{cyan}
        \node[anchor=south east] at (0,0) {$T^f_D$};
        \draw[midarrow] (\shift,\shift) -- ++(\shift,\shift);
        \wt{\shift}{\shift}
        \node[anchor=south] at (\shift,\shift) {$U_p$};
        \draw[midarrow] (1,0) to (0.5,0);
        \wt{0.5}{0}[lime]
        \node[anchor=north] at (0.5,-0.05) {$U^{DC}_v$};
    \end{diagram}
    \\
    \begin{diagram}[0.8][1.2]
        \dobase{0}{0} \drawTket{0}{0}{blue}
        \node[anchor=south east] at (0,0) {$T^b_A$};
    \end{diagram} &= \begin{diagram}[0.8][1.2]
        \dobase{0}{0} \drawTket{0}{0}{blue}
        \node[anchor=south east] at (0,0) {$T^f_A$};
        \draw[midarrow] (\shift,\shift) -- ++(\shift,\shift);
        \wt{\shift}{\shift}
        \node[anchor=south] at (\shift,\shift) {$U_p$};
        \draw[midarrow] (1,0) to (0.5,0);
        \wt{0.5}{0}[green]
        \node[anchor=north] at (0.5,-0.05) {$U^{AB}_v$};
    \end{diagram},
    &
    \begin{diagram}[0.8][1.2]
        \dobase{0}{0} \drawTket{0}{0}{orange}
        \node[anchor=south east] at (0,0) {$T^b_B$};
    \end{diagram} &= \begin{diagram}[0.8][1.2]
        \dobase{0}{0} \drawTket{0}{0}{orange}
        \node[anchor=south east] at (0,0) {$T^f_B$};
        \draw[midarrow] (\shift,\shift) -- ++(\shift,\shift);
        \wt{\shift}{\shift}
        \node[anchor=south] at (\shift,\shift) {$U_p$};
        \draw[midarrow] (-0.5,0) to (-1,0);
        \wt{-0.5}{0}[green]
        \node[anchor=north] at (-0.5,-0.05) {$U^{AB\dagger}_v$};
    \end{diagram}
\end{aligned}
\end{equation}
The tensors $U^{AB}_v, U^{DC}_v$ are odd-parity unitary tensors, 
\begin{subequations}
\begin{align}
    \begin{diagram}[0.8][1.4]
        \dobase{0}{0} 
        \foreach \x in {-0.5,0}
        {\draw[midarrowrev] (\x,0) -- ++(0.5,0);}
        \wt{0}{0}[green]
        \node[anchor=north] at (0,-0.1) {$U^{AB}_v$};
        \node[anchor=east] at (-0.5,0) {$\mu$};
        \node[anchor=west] at (0.5,0) {$\nu$};
    \end{diagram} &= \left(\begin{array}{c|c}
        0 & \mathbbm{1}_{D^{AB}_o} \\ \hline
        \mathbbm{1}_{D^{AB}_e} & 0
    \end{array}\right)_{\mu \nu} \ket{\mu} \bra{\nu},
    \\
    \begin{diagram}[0.8][1.4]
        \dobase{0}{0} 
        \foreach \x in {-0.5,0}
        {\draw[midarrowrev] (\x,0) -- ++(0.5,0);}
        \wt{0}{0}[lime]
        \node[anchor=north] at (0,-0.1) {$U^{DC}_v$};
        \node[anchor=east] at (-0.5,0) {$\mu$};
        \node[anchor=west] at (0.5,0) {$\nu$};
    \end{diagram} &= \left(\begin{array}{c|c}
        0 & \mathbbm{1}_{D^{DC}_o} \\ \hline
        \mathbbm{1}_{D^{DC}_e} & 0
    \end{array}\right)_{\mu \nu} \ket{\mu} \bra{\nu},
\end{align}
\end{subequations}
where $\mathbbm{1}_d$ means a $d$-dimensional identity matrix, $D^{AB/DC}_{e/o}$ is the even/odd dimension of the bond in $\ket{\psi_f}$ on which $U^{AB}_v$/$U^{DC}_v$ resides. 
This transformation preserves the parity of each site tensor in the fTPS. The resulting $\ket{\psi_b}$ in the slave-fermion representation has the same expectation values of physical observables as $\ket{\psi_f}$. 
Therefore, on the square lattice, there is no much difference at the fTPS level (when the unit cell size is $2 \times 2$) between the original electron representation and the fractionalized slave-fermion representation.
% Nevertheless, the holons and spinons in the slave-fermion theory may still be far from the true elementary excitations in a doped Mott insulator, in which more complicated construction like phase string theory \cite{PhysRevB.55.3894} may be necessary.

% \begin{table}
% \caption{Comparison of non-zero tensor elements in the $t$-$J$ model between fermionic and fractionalized Schwinger boson representations.}
% \label{tab:tJ_two_reps}
% \centering
% \begin{tabular}{m{1.8cm}|p{3.2cm}|p{3.2cm}}
%   \hline
%     & hopping ($-t$) & Heisenberg ($J$) \\ 
%   \hline
%   fermion rep. & $H_{1001}^{\textcolor{red}{1001}}=H_{0000}^{\textcolor{red}{1001}}=1$ \newline $H_{0110}^{\textcolor{red}{0110}}=H_{0000}^{\textcolor{red}{0110}}=-1$ &     $H_{1001}^{\textcolor{red}{1111}}=H_{0110}^{\textcolor{red}{1111}}=\frac{1}{2}$ \newline
%     $H_{1100}^{\textcolor{red}{1111}}=H_{0011}^{\textcolor{red}{1111}}=-\frac{1}{2}$ \\ 
%   \hline
%   S. boson rep. & $H_{1001}^{\textcolor{blue}{0110}}=H_{0000}^{\textcolor{blue}{0110}}=1$ \newline $H_{0110}^{\textcolor{blue}{1001}}=H_{0000}^{\textcolor{blue}{1001}}=-1$ &  $H_{1001}^{\textcolor{blue}{0000}}=H_{0110}^{\textcolor{blue}{0000}}=\frac{1}{2}$ \newline
%     $H_{1100}^{\textcolor{blue}{0000}}=H_{0011}^{\textcolor{blue}{0000}}=-\frac{1}{2}$ \\ 
%   \hline
% \end{tabular}
% \end{table}

\newpage
\clearpage
% \onecolumngrid
% \section{More analytical details}

\appendix

\onecolumngrid
\textbf{\centering\large Supplementary materials for "Competing pair density wave orders in the square lattice $t$-$J$ model"}
\vspace{20pt}
\twocolumngrid

\section{Choice of different sub-dimensions and convergence benchmark of VUMPS measurement}

The dimension of a virtual bond of a fermionic tensor reads $D=D_{e}+D_{o}$.
Typically for a given $D$, we could choose to set $D_{e}=D_{o}=D/2$.
However, it is not necessary.
In this section, we test other different choices to see which one leads to the best energy.

\begin{figure}[H]
    \centering
    \includegraphics[width=0.35\textwidth]{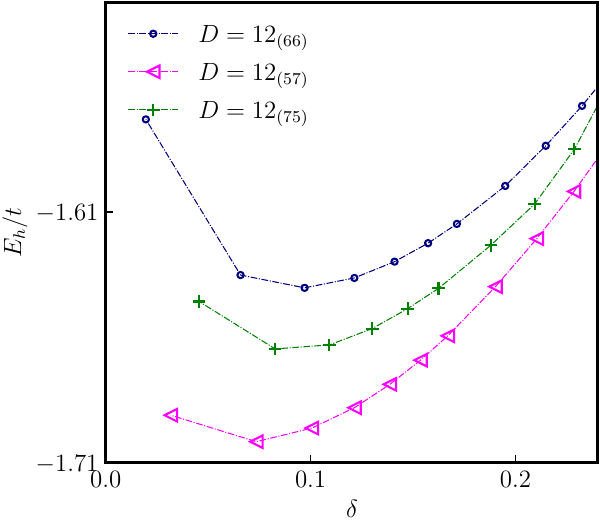}
    \caption{
        Per-hole energies of $d$-wave state in $t$-$J$ model with $t/J = 3.0$.
        $D=12$ and $\chi=48$.
        Sub-dimensions $(D_e, D_o)$ (indicated by the subscripts) are different.
    }
    \label{fig:tJ_diff_splits_hole_ene}
\end{figure}

\begin{figure}[H]
    \centering
    \includegraphics[width=0.48\textwidth]{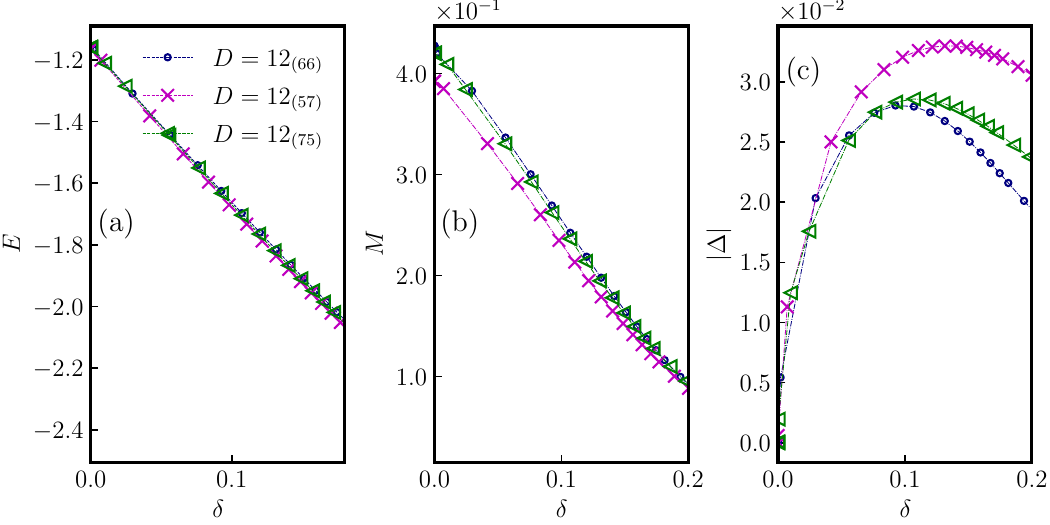}
    \caption{
    Observables from $d$-wave state in $t$-$J$ with with $t/J = 3.0$.
    $D = 12$ and $\chi=48$.
    Subscripts in the bracket denote $(D_e, D_o)$.
    (a) Per-site ground state energy.
    (b) Magnetization.
    (c) Singlet pairing's magnitude.
    }
    \label{fig:tJ_diff_splits_dwave}
\end{figure}

As shown in Fig.~\ref{fig:tJ_diff_splits_hole_ene}, for the example $D=12$, we find that $D=12=5+7$ gives us the best energy rather than the naive $D=6+6$.
In Fig.~\ref{fig:tJ_diff_splits_dwave}, we can see that $D=12=5+7$ not only gives the best energy but produces smaller magnetization and larger superconductivity magnitude at the same time.
Without providing specific justifications, the ideal sub-dimensions are tested and presumed based on the energy for all the data in the main text.

\begin{figure}[H]
    \centering
    \includegraphics[width=0.48\textwidth]{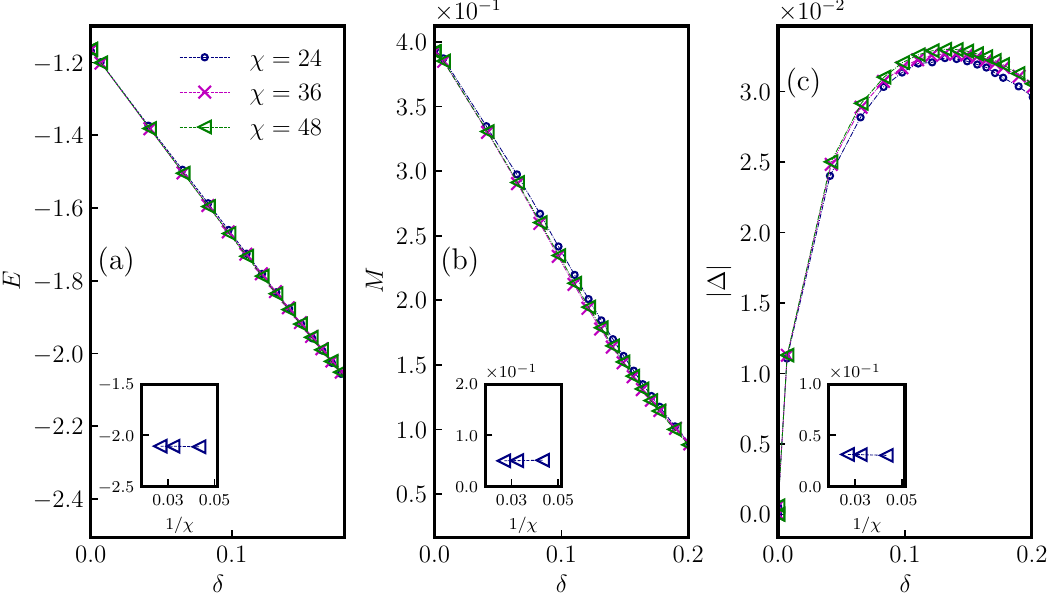}
    \caption{
    Convergence benchmark of VUMPS measurement of $d$-wave state in $t$-$J$ model.
    $D = 12$.
    Boundary MPS is computed with $\chi=2D, 3D, 4D$, respectively.
    (a) Per-site ground state energy.
    (b) Magnetization.
    (c) Singlet pairing's magnitude.
    All illustrative insets are plotted with $\mu=5.0$.}
    \label{fig:tJ_dwave_convergene_benchmark_D12}
\end{figure}

On the other hand, we need to guarantee the convergence for the boundary VUMPS measurement.
For example $D=12$, as shown in Fig.~\ref{fig:tJ_dwave_convergene_benchmark_D12}, it indicates that $\chi=4D$ can lead to a well accepted convergence with respect to these physical observables.
All the data presented in the main text are computed with $\chi\geq{4D}$ starting from $\chi=2D$ to guarantee the convergence.
It should be noted that compared to other projection techniques like the corner transfer matrix renormalization group (CTMRG), VUMPS, being a variational method, usually requires smaller boundary dimensions for convergence.

\begin{figure}[H]
    \centering
    \includegraphics[width=0.45\textwidth]{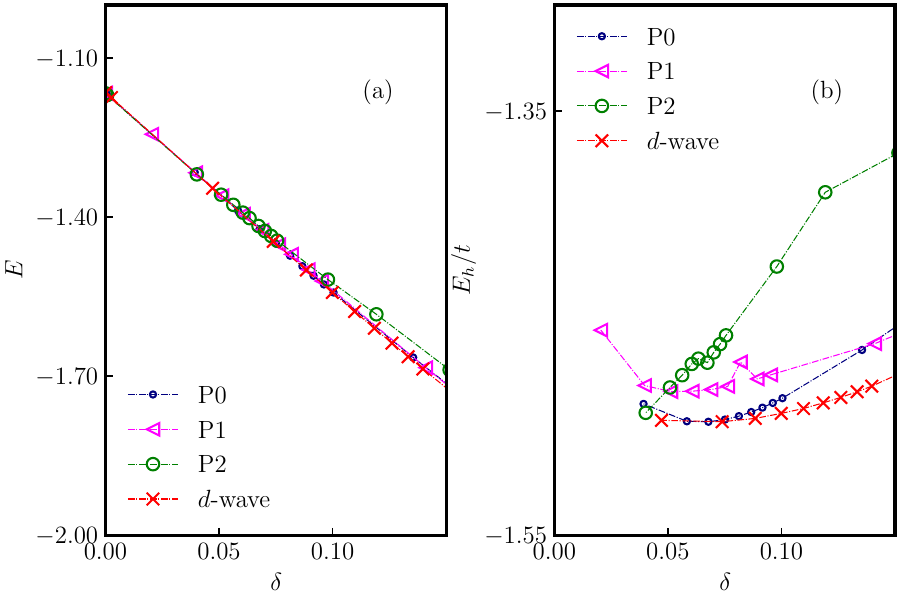}
    \caption{
    (a) Per-site energy $E(\delta)$ of different competing states obtained from $t$-$J$ model with $t/J=2.5$.
    $D=14$ and $\chi=56$.
    (b) Per-hole energy $E_{h}(\delta)$ of these states. 
    }
    \label{fig:tJ_ene_t25}
\end{figure}

%\begin{figure}[H]
%    \centering
%    \includegraphics[width=0.5\textwidth]{tJ_vumps_PDWs_mag_t25.pdf}
%    \caption{
%    Estimation of magnetization $M(\delta)$ as $D \to \infty$ of competing states in $t$-$J$ model from a $1/D$ extrapolation.
%    $t/J=2.5$.
%    (a, b, c, d) correspond to P0, P1, P2 and $d$-wave state, respectively.
%    Each $M(\delta)$ is fitted by a fourth-degree polynomial and plotted as a dashed line with the same color.
%    }
%    \label{fig:tJ_Dscaling_mag_t25}
%\end{figure}

\begin{figure}[H]
    \centering
    \includegraphics[width=0.52\textwidth]{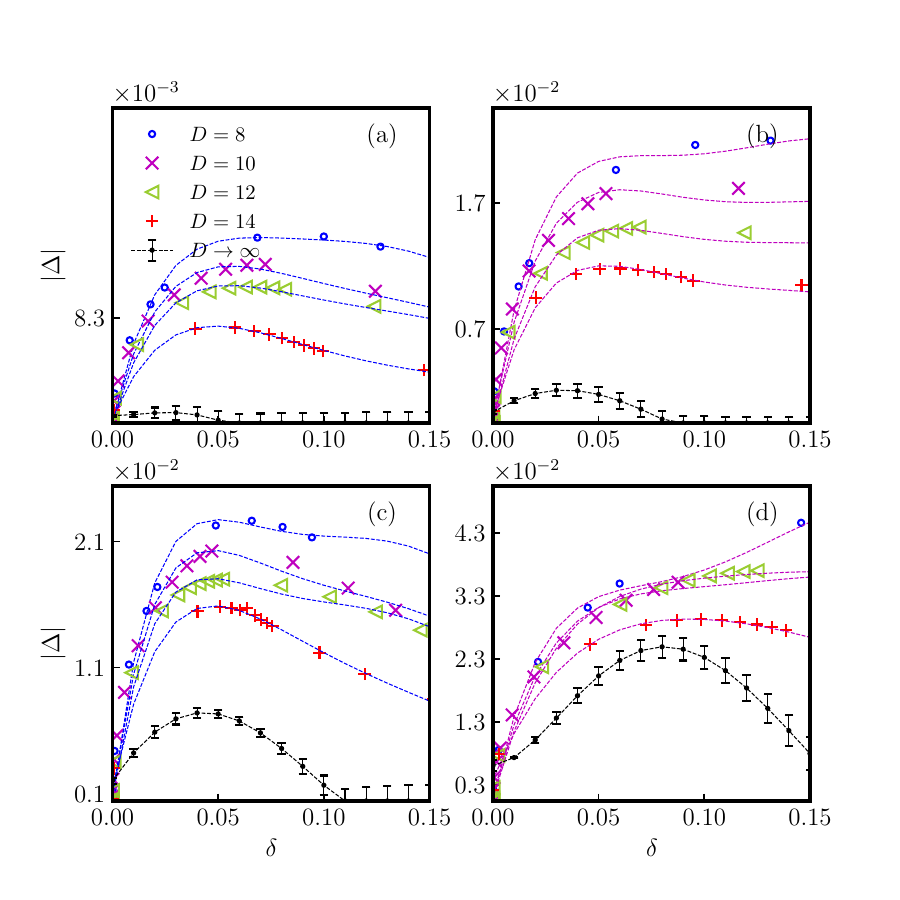}
    \caption{
    Estimation of singlet pairing magnitudes $\vert\Delta\vert(\delta)$ as $D\to\infty$ for competing states in $t$-$J$ model from a $1/D$ extrapolation.
    $t/J=2.5$.
    (a, b, c, d) correspond to P0, P1, P2 and $d$-wave, respectively.
    Each $\vert\Delta\vert(\delta)$ is fitted by a sixth-order polynomial and plotted as a dashed line with the same color.
    }
    \label{fig:tJ_Dscaling_sp_t25}
\end{figure}

\begin{figure}[H]
    \centering
    \includegraphics[width=0.45\textwidth]{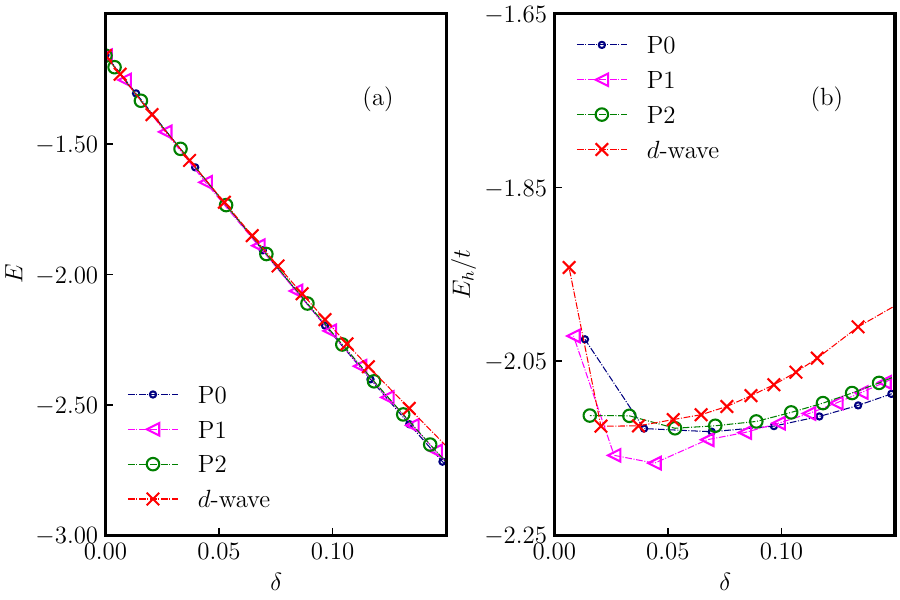}
    \caption{
    (a) Per-site energy $E(\delta)$ of different competing states obtained from $t$-$J$ model with $t/J=5.0$.
    $D=12$ and $\chi=48$.
    (b) Per-hole energy $E_{h}(\delta)$ of these states. 
    }
    \label{fig:tJ_ene_t50}
\end{figure}

\begin{figure}[H]
    \centering
    \includegraphics[width=0.52\textwidth]{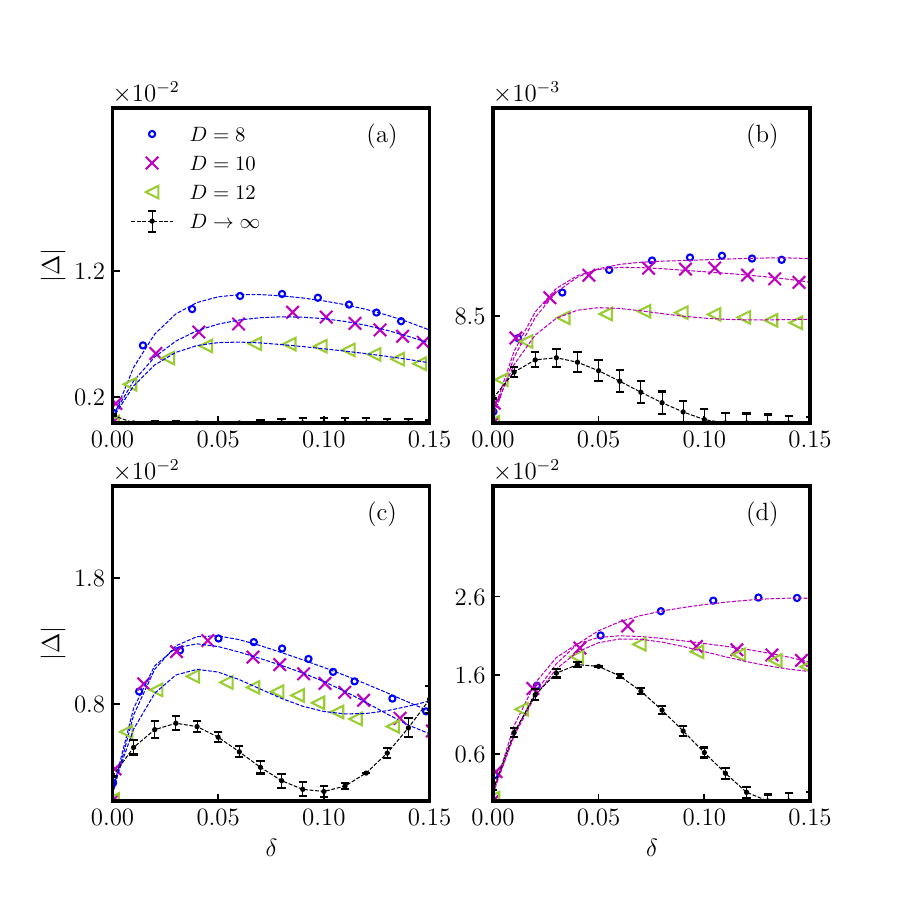}
    \caption{
    Estimation of singlet pairing magnitudes $\vert\Delta\vert(\delta)$ as $D\to\infty$ for competing states in $t$-$J$ model from a $1/D$ extrapolation.
    $t/J=5.0$.
    (a, b, c, d) correspond to P0, P1, P2 and $d$-wave, respectively.
    Each $\vert\Delta\vert(\delta)$ is fitted by a sixth-order polynomial and plotted as a dashed line with the same color.
    }
    \label{fig:tJ_Dscaling_sp_t50}
\end{figure}

%\begin{figure}[H]
%    \centering
%    \includegraphics[width=0.45\textwidth]{tJ_vumps_PDWs_mag_t50.pdf}
%    \caption{
%    Estimation of magnetization $M(\delta)$ as $D \to \infty$ of competing states in $t$-$J$ model from a $1/D$ extrapolation.
%    $t/J=5.0$.
%    (a, b, c, d) correspond to P0, P1, P2 and $d$-wave state, respectively.
%    Each $M(\delta)$ is fitted by a fourth-degree polynomial and plotted as a dashed line with the same color.
    % Insets are extrapolation examples with $\delta=0.125$.
%    }
%    \label{fig:tJ_Dscaling_mag_t50}
%\end{figure}

\section{More data for other ratios $t/J=2.5$ and $t/J=5.0$}

In this section, in addition to $t/J=3.0$ discussed in the main text, we show more data when it comes to $t/J=2.5$ and $t/J=5.0$ since the experimental suggestion is not precisely located at $t/J=3.0$~\cite{RevModPhys.78.17}.
In brief, the conclusions are similar saying that the low-energy subspace of $t$-$J$ model contains several competing real-space Cooper paired competing states and is highly fluctuating.

Fig.~\ref{fig:tJ_ene_t25} and Fig.~\ref{fig:tJ_Dscaling_sp_t25} show data for $t/J=2.5$.
For $D=14$, the extrapolated energy at half-filling is $\tilde{E}_0\approx{-1.166(2)}$, which is about $0.27\%$ higher than the best quantum Monte Carlo result $E_0=-1.1694$ \cite{PhysRevB.56.11678} of Heisenberg model.
We also make extrapolated estimations of singlet pairing's magnitude for these states as $D\to\infty$, which are shown in Fig. \ref{fig:tJ_Dscaling_sp_t25}.
We can see that, similar to $t/J=3.0$, the plaquette P2 state still possesses the strongest pairing amplitude among PDW states in the underdoped regime.
Its energy goes down very quickly particularly for dopings $\delta<0.05$.
Fig.~\ref{fig:tJ_ene_t50} and Fig.~\ref{fig:tJ_Dscaling_sp_t50} show repeated data for $t/J=5.0$.
As shown in Fig.~\ref{fig:tJ_ene_t50}, up to $D=12$, it seems that $d$-wave state is not as good as it for $t/J=3.0$.
Note that $t/J=5.0$ only works as toy model for theoretical test. In general, large $t/J$ ratio will stabilize PDW states, while
$t/J=3.0$ and $t/J=2.5$ which are closer to the real physics in materials will stabilize the usual d-wave state.
% Moreover, we realize that our method works better for smaller $t/J$ ratios than larger ones.
% For example, in Fig.~\ref{fig:tJ_Dscaling_mag_t50} with $t/J=5.0$, the magnetizations obtained from different $D$s are almost identical, which is physically counterintuitive.
% Magnetizations should be suppressed further by larger $D$ as shown in cases of $t/J=2.5$ and $t/J=3.0$.

\section{Possible tendency towards inhomogeneity}

\begin{figure}[H]
    \centering
    \includegraphics[width=0.45\textwidth]{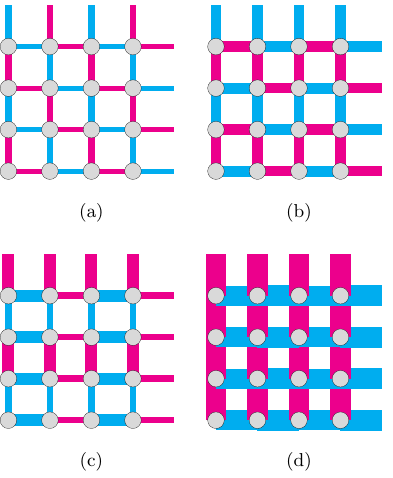}
    \caption{
    Inhomogeneous PDW patterns emerge from uniform ones if bond weights are not averaged.
    (a, b, c, d) correspond to P0, P1, P2 and $d$-wave state, respectively.
    Patterns are plotted with $t/J=3.0$.
    $D=10$ and $\chi=40$.
    Plotted bond width is proportional to the computed $|\Delta|$ around the optimal doping $\delta\approx 12.5\%$.
    }
    \label{fig:tJ_pdw_patterns_inhom}
\end{figure}

\begin{figure}[H]
    \centering
    \includegraphics[width=0.35\textwidth]{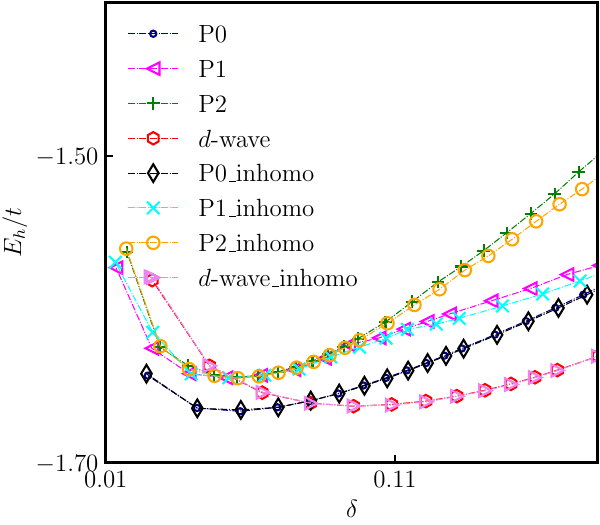}
    \caption{Per-hole energy $E_{h}(\delta)$ for inhomogeneous states in comparison with homogeneous ones.
    $t/J=3.0$.
    $D=10$ and $\chi=40$.
    }
    \label{fig:tJ_hole_ene_t30_inhom}
\end{figure}

In this section, we would like to address the inhomogeneity issue more as we mentioned in the main text.
% present some inhomogeneous properties we observed in the numerical simulations.
We take a test on $t/J=3.0$ and $D=10$.
Starting from converged uniform states shown in Fig.~\ref{fig:tJ_pdw_patterns}, if we skip the weight-averaging during the imaginary time evolution process, we find that these PDW states would possibly evolve to inhomogeneous ones shown in Fig.~\ref{fig:tJ_pdw_patterns_inhom} while $d$-wave does not seem to have such a strong inhomogenuous tendency.
In Fig.~\ref{fig:tJ_hole_ene_t30_inhom}, we can see that the inhomogenuous P1 and P2 states have slightly lower energies than the homogenuous ones.
$d$-wave and P0 seem to have the same energies.
% This implies that homogeneous states could be some kind of semi-stable ones.
It could be an intrinsic feature of high-$T_{c}$ cuprate superconductors rather than the seeming dirt of a cuprate because of its chemical doping.
Inhomogeneity can gradually emerge from a pure $t$-$J$ model.

We would like to argue that such a inhomogeneous tendency is likely one of the causes of long range non-uniform charge orders reported before~\cite{PhysRevB.84.041108}. In this sense, the seemingly complexity of cuprates can be divided into two aspects:
Firstly, the low-energy subspace of $t$-$J$ model contains many superconducting competing states.
They are highly fluctuating and could be superposed.
Secondly, there is a possible additional mechanism to drive the states into non-uniform ones.

\bibliographystyle{apsrev4-2} 
\bibliography{refs}

\end{document}